\documentclass[%
preprint,
nofootinbib,
amsmath,
amssymb,
preprintnumber,
]{revtex4}

\usepackage{graphicx}
\usepackage{dcolumn}
\usepackage{bm}
\usepackage[dvipsnames]{xcolor}

\begin{document}

\preprint{CYCU-HEP-22-07}

\title{Open String Probe in Soft Hair BTZ}%

\author{Chi-Hsien Tai}\thanks{xkp92214@gmail.com}
 \affiliation{Department of Physics and Center for Theoretical Physics,\\
 Chung Yuan Christian University, Taoyuan, Taiwan}
\author{Sayid Mondal}\thanks{sayid.mondal@gmail.com}
 \affiliation{Department of Physics and Center for Theoretical Physics,\\
 Chung Yuan Christian University, Taoyuan, Taiwan}
\author{Wen-Yu Wen}\thanks{wenw@cycu.edu.tw}
 \affiliation{Department of Physics and Center for Theoretical Physics,\\
 Chung Yuan Christian University, Taoyuan, Taiwan}
 \affiliation{Leung Center for Cosmology and Particle Astrophysics, National Taiwan University, Taipei, Taiwan}

\begin{abstract}
We study the boundary CFT under supertranslation in terms of the AdS/CFT correspondence.  In particular, we probe the soft hair BTZ background by several open string configurations as follows.  The Ryu-Takayanaki formula of holographic entanglement remains intact under supertranslation.  The U-shape string probe indicates that the flat part of meson potential is altered by the soft charge.  The trailing string solution shows that the hair parameter plays the role of external field in the drag force.  The study of hanging string in the Brownian motion suggests the transport variables, such as friction coefficient are likely affected by supertranslation.
\end{abstract}


\maketitle


\section{Introduction}
 
The New Massive Gravity (NMG) theory describes a ghost-free, parity-even massive gravity and its action reads:
\begin{equation}
 I = \frac{1}{16\pi G} \int d^3x \sqrt{-g} \big( R-2\lambda - \frac{1}{\mathfrak{m}^2}K \big), \qquad K \equiv R_{\mu\nu}R^{\mu\nu} - \frac{3}{8}R^2.   
\end{equation}
NMG is attractive because at the linearized level it reproduces the unitary Pauli-Fierz action for a massive graviton \cite{Bergshoeff:2009hq}.  While in the trivial limit $\mathfrak{m}^2 \to \infty$, the theory reduces to usual GR, at the special point $\mathfrak{m}^2=\lambda$ it admits a one-parameter hairy BTZ solution \cite{Donnay:2020yxw}
\begin{equation}
    ds^2 = -(\frac{r^2}{L^2}+br-m)dt^2 + (\frac{r^2}{L^2}+br-m)^{-1}dr^2 + r^2d\phi^2.
\end{equation}
The supertranslation symmetry associated with soft hair $b$ is a pure gauge at the asymptotic boundary in the sense that the physical mass remains the same under the shift:
\begin{equation}\label{eqn:gauge_transform}
    \delta r = -r_\epsilon,\qquad \delta b = \frac{2r_\epsilon}{L^2},\qquad \delta m = -br_\epsilon - \frac{r_\epsilon^2}{L^2}.
\end{equation}
Upon gauging away, that is $b\to 0$, one recovers the nonrotating BTZ black hole \cite{Banados:1992wn}.  Though the bulk quantities of hairy black hole are invariant under supertranslation, the corresponding CFT on the boundary may not necessarily respect the same symmetry.  In this letter, we will inspect this discrepancy by probing the boundary field theory with several open string configurations.
The article is organized as follows: In the section II, the Ryu-Takayanaki formula is adopted to study holographic entanglement.  In the section III, we compute the Wilson loop by a U-shape string probe and obtain the meson potential in Coulomb and deconfining phase.  In the section IV, a heavy quark moving in the Quark-Gluon-Plasma is modelled by a trailing string and we compute the drag force.  In the section V, we study the Brownian motion by a  hanging string probe in the rotating background and compute the friction coefficient.

\section{Entanglement}
The soft-haired metric can be recasted into a standard form by shifting $\hat{r}=r+bL^2/2$:
\begin{equation}\label{eqn:shift_metric}
    ds^2 = -(\frac{\hat{r}^2}{L^2}-M)dt^2 + (\frac{\hat{r}^2}{L^2}-M)^{-1}d\hat{r}^2 + (\hat{r}-\frac{bL^2}{2})^2d\phi^2,
\end{equation}
with its physical mass defined at the asymptotic AdS$_3$,
\begin{equation}\label{eqn:hair_mass}
M=\frac{m}{4G}+\frac{b^2L^2}{16G}.
\end{equation}
Consider the one-dimensional boundary is divided into two parts, each of length $\l$ and $2\pi -\l$.  The entanglement between these two parts is given by the Ryu-Takayanaki formula \cite{Ryu:2006bv}:
\begin{equation}
    S = \frac{c}{3}\log(\frac{\beta}{\pi \epsilon} \sinh{\frac{\pi \l}{\beta}}),
\end{equation}
where the inverse of temperature $\beta = \frac{\pi L}{\sqrt{GM}}$.  The central charge $c=\frac{3L}{G}$ is not altered by the supertranslation \cite{Donnay:2020yxw}.  At small $b$ limit,
\begin{equation}\label{eqn:RT_entropy}
    S \simeq \frac{c}{3}\log\frac{\l}{\epsilon} + \frac{cm\l^2}{72L^2}+ \frac{c\l^2}{288}b^2 + \cdots = \frac{c}{3}\log\frac{\l}{\epsilon} + \frac{c\l^2G}{18L^2}M + \cdots,
\end{equation}
here the correction ${\cal O}(b^2)$ can be absorbed into the definition of physical mass $M$ (\ref{eqn:hair_mass}).  The first term in (\ref{eqn:RT_entropy}) is entanglement entropy ($c/3$ times geodesic length) in the absence of black hole, while the second term can also be expressed as $\frac{\l^2M}{6L}$ (by replacing G), which is universal (independent of central charge $c$).  Since the physical mass $M$ is invariant under the gauge transformation (\ref{eqn:gauge_transform}), it is straightforward to conclude that the entanglement entropy is also invariant under supertranslation.

\section{U-shape string}
\begin{figure}
\includegraphics[scale=0.5]{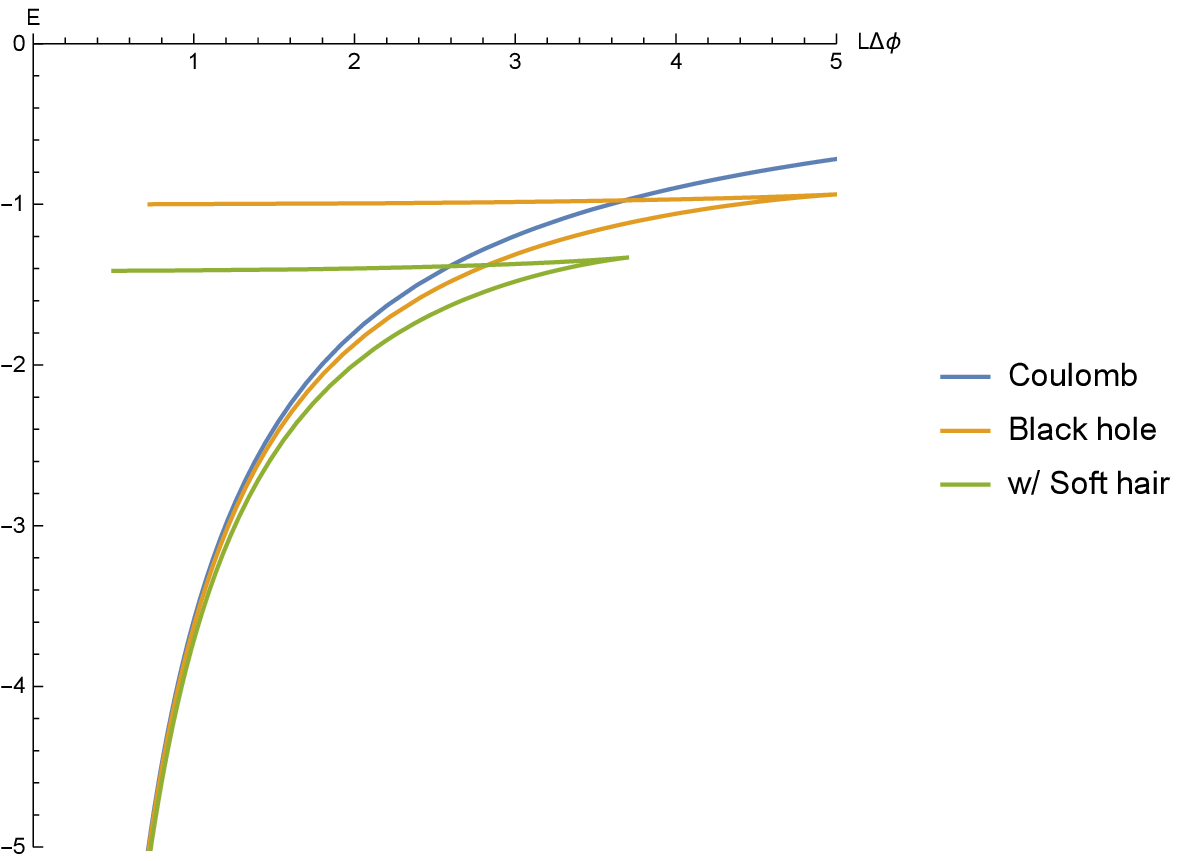}
\includegraphics[scale=0.5]{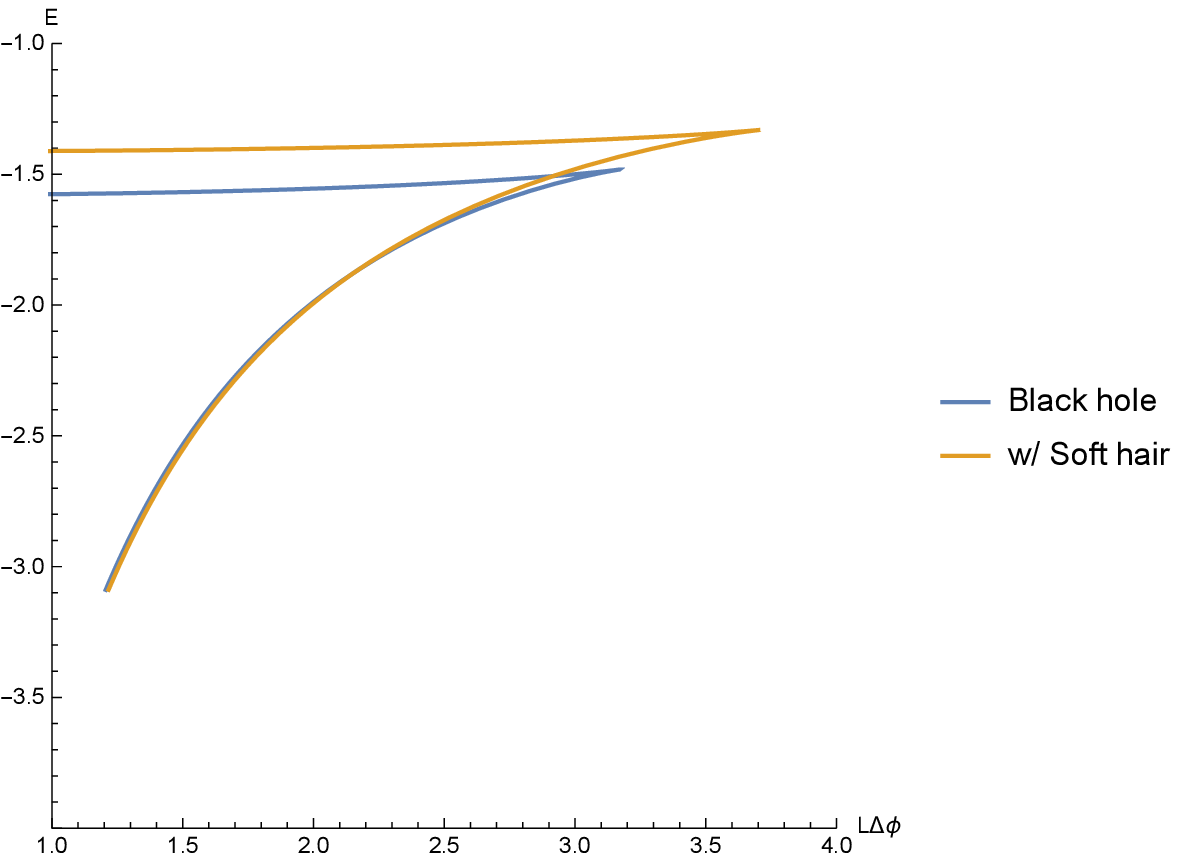}
\caption{\label{fig:potential}(a) To the left, the meson potential $E$ is plotted against the separation $\Delta\Phi$ for chosen parameters $M$ and $b$.  The potential is Coulomb-type in pure AdS.  The pair begins to dissociate at some distance in the black hole background while the potential turns to flat.  The soft charge has the effect to shift the black hole mass as shown in (\ref{eqn:hair_mass}), which corresponds to raise the temperature and make dissociate easier.  (b) To the right we compare the potential given by a pure black hole to another one with soft hair.  By adjusting the pure black hole mass, one may partially fit only the Coulomb part of potential but have large discrepancy for the flat part of potential. 
}
\end{figure}

According to the AdS/CFT dictionary, a long open string having its endpoint at the boundary represents a quark or antiquark, where the string tension contributes to its mass.  A Wilson loop can be constructed by joining a pair of quark and antiquark (meson) by a U-shape string, which is useful to probe the phase of boundary field theory \cite{Maldacena:1998im}.  Computation in pure AdS space shows a generic feature of Coulomb potential and the existence of black hole may dissociate the pair and corresponds to the deconfining phase.  Here we begin with the shifted metric (\ref{eqn:shift_metric}) and embedding ansatz
\begin{equation}
    \tau = t,\qquad \sigma = \hat{r}.
\end{equation}
Let $\phi(\hat{r})$ give the profile of U-shape string and its derivative can be obtained via variation of worldsheet action:
\begin{equation}\label{eqn:phi_prime}
\phi^\prime(\hat{r})=\sqrt{\frac{-g_{tt}(\hat{r}_m)g_{\phi\phi}(\hat{r}_m)}{g_{tt}(\hat{r}) g_{\phi\phi}(\hat{r})\big[g_{tt}(\hat{r}) g_{\phi\phi}(\hat{r})-g_{tt}(\hat{r}_m) g_{\phi\phi}(\hat{r}_m)\big]}}.
\end{equation}
Here $\hat{r}_m$ is the lowest point of U-shape string where $\phi^\prime$ blows up.  The separation is related to $\hat{r}_m$ by a complicated integral:
\begin{equation}
\Delta \phi = \frac{2L}{\hat{r}_m}\int_1^\infty{dy \frac{(1-\hat{b})\sqrt{(1-\hat{M})}}{\sqrt{(y^2-\hat{M})(y-\hat{b})^2\big[(y^2-\hat{M})(y-\hat{b})^2-(1-\hat{M})(1-\hat{b})^2 \big]}}},
\end{equation}
where it is convenient to introduce new variable $y=\frac{\hat{r}}{\hat{r}_m}$ and parameters $\hat{M}=\frac{L^2}{\hat{r}_m^2}M, \quad \hat{b}=\frac{L^2}{2\hat{r}_m}b$.  The integral reduces to the known result $\frac{\Gamma[3/4]}{\Gamma[1/4]}\sqrt{\pi}$ as both $M$ and $b$ vanish.  In generic, $\Delta \phi$ is not simply inversely proportional to $\hat{r}_m$ as in the case of Coulomb potential.
On the other hand, calculation of Wilson loop using solution (\ref{eqn:phi_prime}) gives the finite meson potential:
\begin{equation}
 E = \frac{\hat{r}_m}{\pi\alpha^\prime}\bigg\{\int_1^\infty{dy\bigg[\sqrt{\frac{(y^2-\hat{M})(y-\hat{b})^2}{(y^2-\hat{M})(y-\hat{b})^2-(1-\hat{M})(1-\hat{b})^2}}-1\bigg]}-1\bigg\}.   
\end{equation}
This integral gives a negative finite value $-0.5991$ as both $M$ and $b$ vanish, indicating an attractive force.  In generic, however, $E$ is not simply proportional to $\hat{r}_m$.  In the Figure \ref{fig:potential} (a), we plot $E$ against $\Delta\phi$ for pure AdS, BTZ black hole and that with soft hair.  The meson dissociates at shorted distance in the hairy BTZ than that without hair, in agreement with the definition (\ref{eqn:hair_mass}) where physical black hole mass is increased by the hair.  The potential is deviated from the Coulomb-type as $\hat{r}_m$ approaches the horizon and suddenly becomes flat as soon as it falls behind (second-order phase transition).  Though it appears that by using (\ref{eqn:gauge_transform}) one may gauge away the soft charge as shown in the Figure \ref{fig:potential} (b), the shift of $\hat{r}_m$ has the effect of shifting flat potential and changing the behavior near phase transition.  For example, the dissociation distance increases by soft hair charge (meson has larger size) even though both have similar Coulomb-like potential.

\section{Trailing string}
The trailing string was used to model the drag force exerting on a quark moving uniformly in the quark-gluon-plasma \cite{Gubser:2006bz}.  Here we assume the end point of an open string moving at uniform speed $v$ and the other end is trailing behind into the bulk.
\begin{equation}\label{eqn:uniform_motion}
    \phi = \frac{v}{L}t + \xi(r)
\end{equation}
where $\xi(r)$ is the string profile in the comoving frame.  The embedding ansatz
\begin{equation}
    \tau = t, \qquad \sigma = r.
\end{equation}
The induced world-sheet horizon locates at $r=r_s$, such that
\begin{equation}
    g_{tt}(r_s) + g_{\phi\phi}(r_s)\frac{v^2}{L^2}=0.
\end{equation}
One obtains
\begin{equation}\label{eqn:worldsheet_horizon}
    r_s = (-L^2b \pm \sqrt{L^4b^2+4L^2 m (1-v^2)})/2(1-v^2)
\end{equation}
The drag force reads
\begin{eqnarray}\label{eqn:drag_force}
    F_d &=&-\frac{1}{2\pi \alpha^\prime}\sqrt{-g_{tt}(r_s)g_{\phi\phi}(r_s)}\nonumber\\
    &\simeq& -\frac{m L}{2\pi\alpha^\prime}\frac{v}{1-v^2} + \frac{\sqrt{m}L^2}{2\pi \alpha^\prime} \frac{v}{1-v^2}\frac{b}{\sqrt{1-v^2}} + {\cal O}(b^2)
\end{eqnarray}
We remark that the force recovers the known result \cite{Bena:2019wcn} at the limit $b=0$.  The gauge transformation (\ref{eqn:gauge_transform}) is no longer valid for this dynamic string.  Nevertheless, a new transformation dressed by the Lorentz factor as follows:
\begin{equation}\label{eqn:dress_transform}
    \delta r = -r_\epsilon,\qquad \delta b = \frac{2r_\epsilon}{L^2}(1-v^2),\qquad \delta m = -br_\epsilon - \frac{r_\epsilon^2}{L^2}(1-v^2).
\end{equation}
can still make the world-sheet horizon (\ref{eqn:worldsheet_horizon}) invariant.  This is possible because the Killing vector associated with supertranslation $\partial_r$ commutes with the Lorentz generator $\phi\partial_t - t\partial_{\phi}$.  The second term in (\ref{eqn:drag_force}) could be regarded as a moving charged quark exerted by an external force due to boundary field $b$ \footnote{Here we might regard this external force as the magnetic force simply because it takes a similar form $F \propto vB$ in the Lorentz force.  However, we are also reminded that in the two-dimensional boundary only field component $F_{t\phi}$ exists, therefore this external force could also be regarded as an electric one.}.

\section{Hanging string in Brownian motion}
In order to have an open string hanging straight into the black hole, that is to set $\xi(r)$ a constant function in the ansatz (\ref{eqn:uniform_motion}), we need to consider a rotating reference frame such that the viscous drag force is balanced with the frame dragging force.  The rotating hairy BTZ solution reads \cite{Donnay:2020yxw}: 
\begin{eqnarray}
ds^2 &=& -N^2(r) F(r) dt^2 + \frac{dr^2}{F(r)} + (r^2+r_0^2)(N^{\phi}(r) dt+ d\phi)^2,\nonumber\\
F(r)&=&\frac{r^2}{L^2}+\frac{(\eta+1)b}{2}r - m \eta + \frac{b^2 L^2 (1-\eta)^2}{16}, \nonumber\\
N^2(r)&=& \frac{\big(4r^2+b^2L^2(1-\eta)\big)^2}{16r^2+L^2(1-\eta)(8m+b^2L^2(1-\eta))}\nonumber\\
N^{\phi}(r)&=& \frac{a(br-m)}{2(r^2+r_0^2)} \nonumber\\
r_0^2&=&\frac{L^2(1-\eta)(8m+b^2L^2(1-\eta))}{16}
\end{eqnarray}
where $a$ is the rotating parameter such that $|a|\le L$ is satisfied and $\eta = \sqrt{1-a^2/L^2}$.

The fluctuation of string, regarding as the Brownian motion of an external quark (string end point), satisfies the Langevin equation \cite{NataAtmaja:2013jxi}:
\begin{eqnarray}
&&m_{q}\ddot{X}+\mu \dot{X} = \xi(t),\nonumber\\
&&<\xi(t)\xi(t')>=2 T\mu \delta(t-t')
\end{eqnarray}
where the effective quark mass $m_q$ and friction coefficient $\mu$ read
\begin{equation}
    m_q = \frac{r_c}{2\pi\alpha'}(1-v^2)^{-3/2}, \qquad \mu = \frac{r_+^2+r_0^2}{2\pi\alpha'}
\end{equation}
here the quark is placed on the UV cutoff brane at $r=r_c$, and $r_+$ is the world-sheet horizon which satisfies $N^{\phi}(r_+)=-v/L$.  A gauge transformation similar to (\ref{eqn:gauge_transform}) to keep $F(r)$ invariant can be found:
\begin{eqnarray}\label{eqn:guage_rotation}
\delta r &=& -r_\epsilon ,\qquad \delta b = \frac{4r_\epsilon}{(\eta+1)L^2}, \nonumber\\
\eta \delta m &=& -\frac{r_\epsilon^2}{L^2} -\frac{\eta+1}{2}br_\epsilon + \frac{(\eta-1)^2}{\eta+1}\frac{br_\epsilon}{2}+\frac{(\eta-1)^2}{(\eta+1)^2}\frac{r_\epsilon^2}{L^2}
\end{eqnarray}
We remark that above transformation reduces to (\ref{eqn:gauge_transform}) at static limit $\eta\to 1$.  Upon transformation (\ref{eqn:guage_rotation}), the effective mass and friction coefficient are shifted as follows:
\begin{eqnarray}
     \delta m_q &=& -\frac{r_\epsilon}{2\pi\alpha'}(1-v^2)^{-3/2}, \nonumber\\
     \delta \mu &=& -\frac{aL}{4\pi\alpha'v}\bigg\{ \frac{4r_\epsilon r_+}{(\eta+1)L^2}+\frac{4r_\epsilon^2}{(\eta+1)^2L^2}+ \frac{1-\eta}{1+\eta}b r_\epsilon\bigg\}.
\end{eqnarray}
Despite those finite term with $r_\epsilon$, the last term in $\delta \mu$ which contains soft charge $b$ does not go away for rotating black hole ($\eta < 1$).  This implies the transport variables in the boundary theory, such as friction coefficient, are likely to be gauge dependent. 

\section{Conclusion}
In this letter, we probed the soft hair BTZ background by several open string configurations.  We adopted the Ryu-Takayanaki formula to compute the entanglement between two divided regions in boundary CFT and found it remains intact under supertranslation.  The U-shape string probe was used to compute time-like Wilson loop formed by a pair of quark and antiquark.  That the meson potential changes from Coulomb-type to flat as meson  inter-distance increases indicates a second-order phase transition to deconfinement.  While the effect of soft charge to the Coulomb part of potential can be cancelled by adjusting the BTZ mass, the flat part of potential fails to do so.  The trailing string solution was used to model the drag force acting on a heavy quark while moving in the QGP-like medium.  Our result shows that the hair parameter plays the role of external field in the drag force.  Nevertheless the formula of drag force remains invariant if the soft gauge transformation were twisted by some Lorentz factor.   At last, a hanging string in the rotating BTZ was used to study the Brownian motion of boundary quarks.  Though a new soft gauge transformation specific to the rotating BTZ was found, the friction coefficient is not invariant under such transformation.  Part of the reason could be that while (\ref{eqn:guage_rotation}) keeps $F(r)$ invariant but not $N^{\phi}(r)$.  A rotating coordinate is not like an inertial frame to respect Lorentz symmetry, so it is more subtle or even impossible to find a compatible twist such as (\ref{eqn:dress_transform}) in a rotating background.

\begin{acknowledgments}
The authors are grateful to Prof. Pei-Ming Ho, Prof. Hsien-chung Kao and Dr. Shingo Takeuchi for useful discussion.  Some parts of this work were reported in the 2022 Annual Meeting of the Physical Society of Taiwan.  This work is supported in part by the Taiwan's Ministry of Science and Technology (109-2112-M-033-005-MY3) and the National Center for Theoretical Sciences (NCTS).

\end{acknowledgments}

\end{document}